\documentclass[twocolumn,showkeys,preprintnumbers,amsmath,amssymb]{revtex4}
\usepackage{graphicx}% Include figure files
\usepackage{dcolumn}% Align table columns on decimal point
\usepackage{bm}% bold math
\usepackage{amssymb}
\newcommand{\zemax}{\emph{ZEMAX}\circledR}
\begin{document}

\preprint{Submitted paper}

\title{An original image slicer \\ designed for Integral Field Spectroscopy with NIRSpec/JSWT}% Force line breaks with \\

\author{S. Viv\`es}
\email{Sebastien.Vives@oamp.fr}
\author{E. Prieto}%
% \email{Eric.Prieto@oamp.fr}
\affiliation{%
Laboratoire d'Astrophysique de Marseille,\\ Site des Olives BP8-13376, Marseille, France \\}%

\date{\today}% It is always \today, today,

\begin{abstract}
Integral Field Spectroscopy (IFS) provides a spectrum simultaneously for each spatial sample of an extended, two-dimensional field. It consists of an Integral Field Unit (IFU) which slices and re-arranges the initial field along the entrance slit of a spectrograph. This article presents an original design of IFU based on the advanced image slicer concept~\cite{Content1997}. To reduce optical aberrations, pupil and slit mirrors are disposed in a fan-shaped configuration that means that angles between incident and reflected beams on each elements are minimized. The fan-shaped image slicer improves image quality in terms of wavefront error by a factor~2 comparing with classical image slicer and, furthermore it guaranties a negligible level of differential aberration in the field. As an exemple, we are presenting the design LAM used for its proposal at the NIRSPEC/IFU invitation of tender.
\end{abstract}

\pacs{Valid PACS appear here}% PACS, the Physics and Astronomy
%Use showkeys class option if keyword
\maketitle

%%%%%%%%%%%%%%%%%%%%%%%%%%%%%%%%%%%%%%%%%%%%%%%%%%%%%%%%%%%%%%%%
\section{ \label{sect:intro} Introduction} 
Integral Field Spectroscopy (IFS) provides a spectrum simultaneously for each spatial sample of an extended, two-dimensional field. Basically, an IFS is located in the focal plane of a telescope and is composed by an Integral Field Unit (IFU) and a spectrograph. The IFU acts as a coupler between the telescope and the spectrograph by reformatting optically a rectangular field into a quasi-continuous pseudo-slit located at the entrance focal plane of the spectrograph. Therefore, the light from each pseudo-slit is dispersed to form spectra on the detector and a spectrum can be obtained simultaneously for each spatial sample within the IFU field. 

The IFU contains two main optical sub-systems: the fore-optics and the image slicer. The fore-optics introduces an anamorphic magnification of the field with an aspect ratio of 1$\times$2 onto the set of slicer mirrors optical surfaces. In such way each spatial element of resolution forms a 1$\times$2~pixels image on the detector (i.e. the width of each slice corresponds to 2 pixels), which ensures correct sampling in the dispersion direction (perpendicular to the slices) and prevents under-sampling the spectra. This anamorphism can be avoided if under-sampling spectra is acceptable by science (for example, the SNAP~\cite{Ealet2003} project) or if a spectral dithering mechanism is included in the spectrograph in order to recover for the under-sampling. The image slicer optically divides the anamorphic (or not) two-dimensional field into a large number of contiguous narrow sub-images which are re-arranged along a one-dimensional slit at the entrance focal plane of the spectrograph. 

\begin{figure}
   \begin{center}
   \begin{tabular}{c}
   \includegraphics[width=7cm]{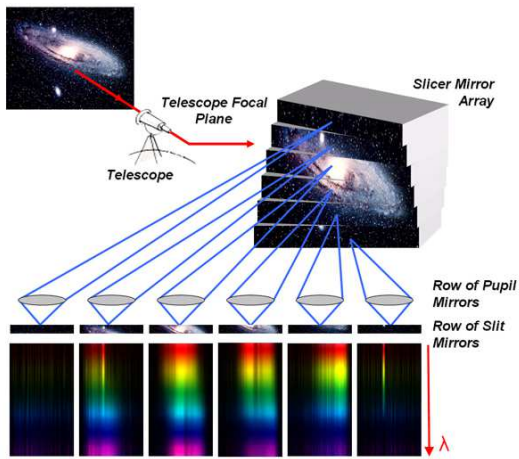}
   \end{tabular}
   \end{center}
   \caption
   { \label{fig:principle} 
The principle of an image slicer. The slicer mirror array, located at the image plane of the telescope, divides the entrance field of view (FOV) and re-mages the telescope exit pupils along a line on the pupil mirrors. Each pupil mirror then re-images its corresponding slice of the entrance FOV on its corresponding slit mirror located at the spectrograph's focal plane (slit plane). The reformatted FOV acts as the entrance slit in the spectrograph where all the slices are aligned as a pseudo long slit. }
   \end{figure}

An image slicer is usually composed of a slicer mirror array located at the image plane of the telescope and associated with a row of pupil mirrors and a row of slit mirrors. The slicer mirror array is constituted of a stack of several thin spherical mirrors (called "slices") which "slice" the anamorphic field and form real images of the telescope pupil on the pupil mirrors. The pupil mirrors are disposed along a row parallel to the spatial direction. Each pupil mirror then re-images its corresponding slice of the anamorphic field on its corresponding slit mirror located at the spectrograph's focal plane (slit plane). The slit mirrors are also diposed along a row parallel to the spatial direction. Finally, each slit mirror which acts as a field lenses, re-images the telescope pupil (pupil mirrors) onto the entrance pupil of the spectrograph. The principle of an image slicer is presented in Fig.~\ref{fig:principle}.

%%%%% FIGURE 1 %%%%%
%%%%%  

In order to improve image quality and/or reduce costs of image slicer, several adaptations have been recently developed:
\begin{itemize}
	\item \textbf{Catadioptric image slicer} where pupil and slit mirrors are replaced by dioptric elements~\cite{Henault2004}. This allows to improve both image performances and costs while increasing the complexity of the opto-mechanical interface in cryogenic environment. Furthermore, dioptric elements present chromatic aberrations and result in a complex arrangement of pupil and slit mirrors since they are close together.
	\item \textbf{"Staggered" image slicer} where pupil mirrors are staggered in two rows instead of a single row~\cite{Henault2004b}. This allows to place the pupil mirrors twice as far away while mainting the slit location so that the largest off-axis angles are reduced by a factor of two and then to improve image quality.
	\item \textbf{Image slicer using a flat facet slicer mirror array}. This image slicer~\cite{Tecza2003} needs an additional spherical or cylindrical field lens located very close the slicer stack. A first look could conduct to think that this configuration optimally reduces cost manufacturing. But nevertheless, the fore-optics has to be more complex (re-imaging the pupil to a precise position after the slicer mirror array), and the progress in glass spherical slices manufacturing process permits to keep the cost differential small for a good system benefit. Furhtermore, such a configuration has the drawback of slightly decreasing the instrument throughput since an additional component is introduced in the optical layout. 
	\item \textbf{Concentric image slicer} where the row of pupil mirrors, the row of slit mirrors and the collimator are disposed along concentric circles centered on the slicer mirror array~\cite{Prieto2004,Dopita2004}. This configuration preserves aberrations in the field of view since angles are identical between each elements of each sub-slit channel. Thus this configuration is well adapted to diffraction limited instruments.
\end{itemize}

It is in the context of improving performances of such a complex system that we propose an original concept of image slicer called "Fan-shaped". As an exemple, we are presenting the design LAM used for its proposal at the NIRSPEC/IFU invitation of tender~\cite{Prieto2003}. The fan-shaped image slicer is described in section~\ref{sect:fanshaped}. Section~\ref{sect:NIRSPEC} is devoted to the description of the whole IFU designed for the NIRSpec/JWST instrument and its performances. Section~\ref{sect:comparison} compares performances of the fan-shaped image slicer with a classical image slicer design.

%We present here the optical design of this image slicer developed in the framework of the JWST/NIRSpec proposal and we compare its performances with a "classical" image slicer. 

%%%%%%%%%%%%%%%%%%%%%%%%%%%%%%%%%%%%%%%%%%%%%%%%%%%%%%%%%%%%%%%%
\section{\label{sect:fanshaped} Principles of the fan-shaped Image Slicer} 

Our purpose was to design an image slicer compliant with the required performances described in Tab.~\ref{tab:requirements}. In particular, the IFU shall have very high performances in terms of image quality and telecentricity while fitting within the small envelope and operating at cryogenic temperature.

As described before, previous designs use diotric elements (micro-lenses) or complex rows of pupil mirrors ("staggered" design) to improve image quality. Because, dioptric elements have a limited bandpass, are not well adapted to cryogenic environment and present chromatic aberrations, the design employs all reflective optics. Each surface of the image slicer are spherical in shape in order to facilitate manufacturability, assembly and alignement of the overall optical system.

Since the main source of aberrations comes from off-axis optics, angles between the incident and the reflected beams must be minimized on each surface in both spectral and spatial directions. 

To limit angles in the spectral direction (perpendicular to the stack slicer), the complete image slicer is slightly tilted until the slicer stack almost sends the light back on itself in this direction. To limit angles in the spatial direction (parallel to the stack slicer), each pupil mirror is located such as the incident chief ray is almost parallel with its normal at the vertex. The pupil mirrors are disposed in a fan-shaped configuration which is coincident with the exit pupil images locations defined by the slicer mirror array. Finally, by tilting each slit mirror it is possible to limit angles between the incident and the reflected beams on each slit mirror. The beam becomes parallel to the main optical axis after the slit mirrors (see Fig.~\ref{fig:FAN}).

The image on the slit is greatly improved, because each channel (composed by a slice and its corresponding pupil and slit mirrors) is now almost on-axis in both directions. Performances are compared with a classical IFU design in the section~\ref{sect:comparison}.

%%%%% FIGURE 2 %%%%%
  \begin{figure*}
   \begin{center}
   \begin{tabular}{c}
   \includegraphics[width=12cm]{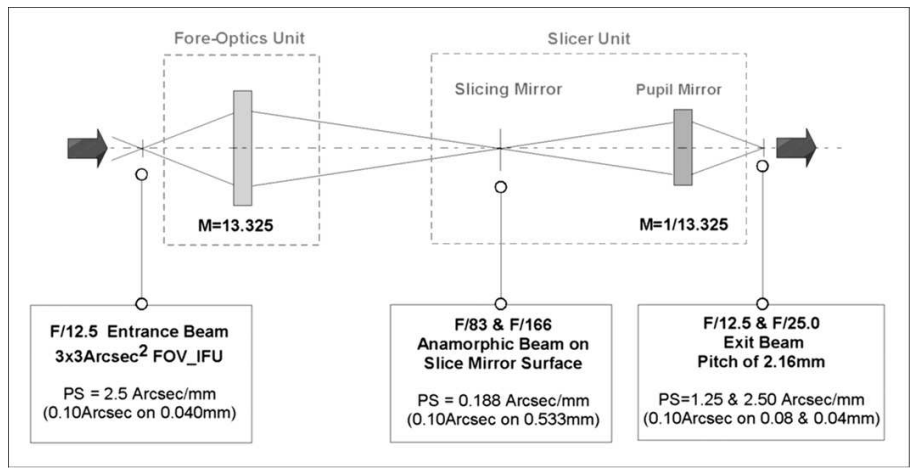}
   \end{tabular}
   \end{center}
   \caption 
   { \label{fig:general} 
The main units of IFU design to produce the required exit beam feeding the spectrograph with appropriate plate scale (PS) and magnification (M).}
   \end{figure*} 
%%%%%  

%%%%%%%%%%%%%%%%%%%%%%%%%%%%%%%%%%%%%%%%%%%%%%%%%%%%%%%%%%%%%%%%
\section{\label{sect:NIRSPEC} Application to NIRSpec/JWST} 
The original concept of "fan-shaped" image slicer described above was successfully appplied to the integral field unit (IFU) of the NIRSpec/JWST instrument. The Near-Infrared Spectrograph (NIRSpec) for JWST is a multi-object spectrograph covering the wavelength range from 0.7~$\mu$m to 5.0~$\mu$m and operating in three distinct modes: multi-object spectrograph mode (MOS), integral field spectrograph mode (IFS) and a canonical long-slit spectrograph mode (LSS). The IFU operates in the IFS mode and reformats a tiny portion NIRSpec field of view in a long entrance slit for the spectrograph. The main IFU requirements were derived from the \textit{NIRSpec requirements specification} document and are listed in Tab.~\ref{tab:requirements}. Every development on the IFU was guided by the lowest possible impact on the spectrograph and other modes of NIRSpec. Thus, the IFU should be integrated apart from the instrument and plug-in during the integration phase of NIRSpec.

%%%%%%%%%%%%%%%%%%%%%%%%%%%%%%%%%%%%%%%%%%%%%%%%%%%%%%%%%%%%%%%%
\subsection{IFU optical design} \label{sect:design}
\begin{figure*}
   \begin{center}
   \begin{tabular}{c}
   \includegraphics[width=12cm]{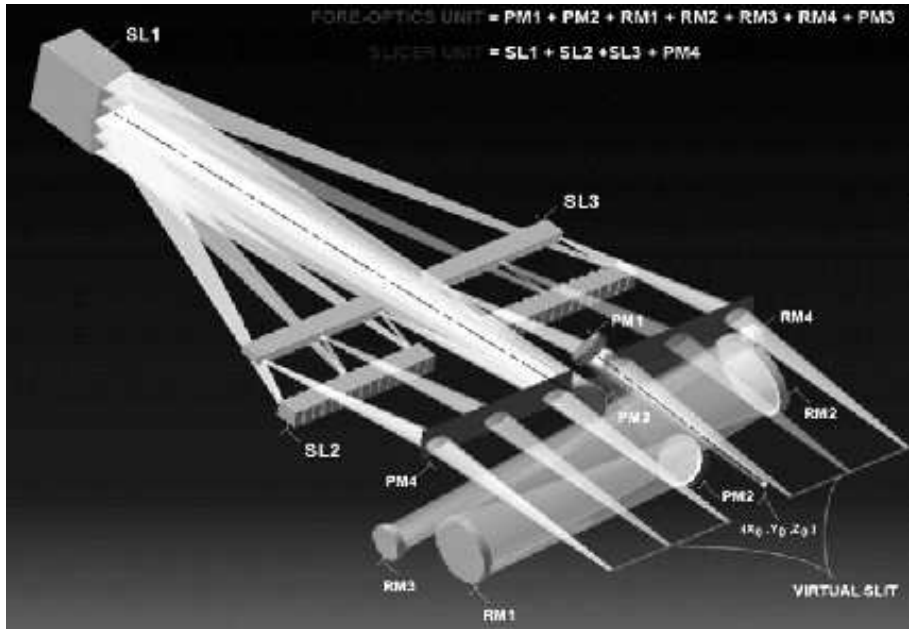}
   \end{tabular}
   \end{center}
   \caption 
   { \label{fig:IFU} 
Overall layout for the complete IFU (fore-optics and image slicer units). The white light beam is the virtual exit beam defining the virtual slit. }
   \end{figure*}

The total spectral range of the IFU (0.7 to 5~$\mu$m) prevents for inserting lenses or dioptric elements within the overall optical layout. An all-reflective design takes advantage of its facility to adapt to a cryogenic environment and presents an higher throughput. All the optical components are made of Zerodur ensuring both preservation of the optical properties of the IFU at operating temperature (35-40~K) and the manufacturability of the IFU (see below). All optical surfaces are gold-coated to maximise their reflectivity in the (near) infrared. The IFU design comprises two units, the fore-optics and the image slicer, as shown schematically in Fig.~\ref{fig:general}. Figure~\ref{fig:IFU} shows the overall layout for the fore-optics and image slicer unit.

\begin{figure}
   \begin{center}
   \begin{tabular}{c}
   \includegraphics[width=7cm]{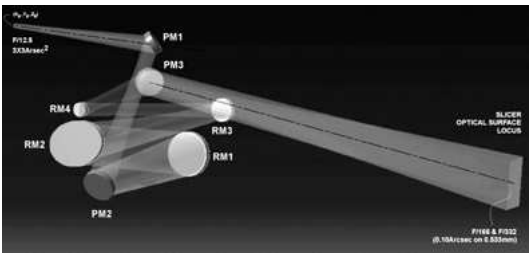}
   \end{tabular}
   \end{center}
   \caption 
   { \label{fig:FO} 
The fore-optics composed of three flat pick-off mirrors (PM1, PM2, PM3) and four powered re-imaging mirrors (RM1, RM2, RM3 and RM4). The optical reference axis is defined by (X$_0$, Y$_0$, Z$_0$) on the centre of the input field of view (3$\times$3~arcsec$^2$). }
   \end{figure}

The fore-optics unit is composed of three flat pick-off mirrors (PM1, PM2, PM3) and four re-imaging mirrors (RM1, RM2, RM3 and RM4) as shown in Fig.~\ref{fig:FO}. The pick-off mirrors are used to capture the input beam (PM1, PM2) and redirect it to the image slicer unit (PM3). The fore-optics unit, used to re-image and magnify the F/12.5 entrance field onto the slicer mirror array, is disposed orthogonally to the main optical axis to fit with the envelope dimension. The re-imaging mirrors RM1, RM2, RM3 are cylindrical mirrors and RM4 is a spherical mirror. The fore-optics seems to be complex by using four mirrors but this choice preserves the manufacturability and allows to reach high surface quality of each component while reducing costs.

%%%%% FIGURE 3 %%%%%
%%%%%  

The slicer unit is composed of the slicer mirror array (SL1), the row of pupil mirrors (SL2) and the row slit mirrors (SL3). To fit within the envelope while reaching the required optical performances, the "fan-shaped" image slicer configuration, described above, was applied. The slicer mirror array is made from a stack of 30 thin mirrors with a typical optical aperture of 10~mm~$\times$~0.55~mm and depth of 10~mm. The pupil image once defined by the fore-optics is placed by each slicer mirror in an optimum position defining a line of discrete sub-pupils images. Each element of the slicer mirror array is spherical and has discrete tilts and curvature. The pupil mirrors unit is made from two sets of 15 mirrors disposed on both sides of the beam coming from the fore-optics (see Fig.~\ref{fig:IFU}). Each pupil mirror is coincident with the line of exit pupil image defined by the corresponding slicer mirror. They re-image each specific slice mirror onto another line configuration defining the locus for each slit mirror. 

%%%%% FIGURE 4 %%%%%
%%%%%  

Some words are needed to understand why we adopt glass standard polished optics instead of monolithic Aluminium diamond turned optics. Maximum throughput means accurate roughness only available using classical polishing technics. Classicaly a roughnes of 0.7~nm can be reached on plane, spherical and cylindrical surfaces (with a surface form of $\lambda/100$) which results in a global transmission of 80\% at 0.7~$\mu$m. Here we considered a surface roughness of 2~nm for each optical component (which is a low level of roughness very standard polishing) to reach an overall throughput better than 65\%. In term of comparison a roughness of 10~nm (which is the standard roughness for surfaces manufactured by diamond turning) on the aluminium results in an overall throughput of 50\%. A critical point about slicer stack manufacturing is the thickness of edges and the fragments size on each sub-mirror element. Using classical polishing technics, edges of 1~$\mu$m with fragments less than 5~$\mu$m are classically reached. It is clear from such estimations the benefit of standard glass manufacturing over the diamond turning manufacturing on the Aluminium. Furhtermore, diamond turning process cannot meet a monolithic system array with individual radius of curvature for each optical sub-aperture.

%%%%%%%%%%%%%%%%%%%%%%%%%%%%%%%%%%%%%%%%%%%%%%%%%%%%%%%%%%%%%%%%
\subsection{\label{sect:performances} IFU Performances}  
Analysis of the design show that the system is compliant with all optical requirements listed in Tab.~\ref{tab:requirements} including tolerancing and margin (calculated after 10000~trials making use of \zemax\ Monte Carlo algorithm capability). We detailled, hereafter, performances in terms of wavefront error (WFE), throughput and telecentricity of the complete IFU. Wavefront error is also discussed in section~\ref{sect:comparison} for the fan-shaped image slicer only.

The tolerancing analysis results show that the overall theoretical WFE shall be less than 55~nm to keep the overall toleranced WFE inside of specifications (100~nm) considering optical manufacturing defects and aligment errors. The mean theoretical WFE calculated over all sub-slit is about 47~nm while considering a combinaison of all optical manufacturing defects and aligment errors, the WFE is about 82~nm in the worst case. This is compliant with the requirement.

Analysis of the system throughput results in an average optical throughput greater than 65\% for any wavelength in the operating range and any position within the FOV. This calculation considers the complete system with gold coated 2~nm roughness surfaces and includes diffraction losses. As mentionned above, glass polishing technics classically reachs a surface roughness of 0.7~nm which results in an overall IFU throughput greater than 80\% at 0.7~$\mu$m (not required here).

\begin{figure}
   \begin{center}
   \begin{tabular}{c}
   \includegraphics[width=6cm]{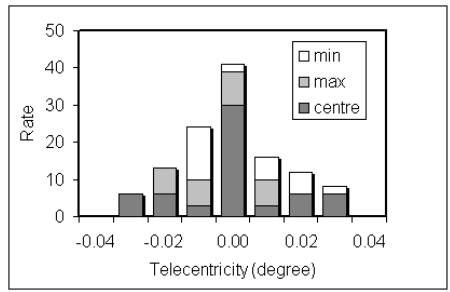}
   \end{tabular}
   \end{center}
   \caption 
   { \label{fig:telecentricity} 
Distribution of the theoretical telecentricity (in degree) considering 3~specific fields over all channels: the centre (dark) and the 2~extremes (white and grey). }
   \end{figure}

The telecentricity requirements corresponds to the position of the pupil on the exit pupil of the spectrograph. The sampling, derived from both the expected final quality of the Point Spread Function (PSF) and the optimized signal to noise ratio on the detector, imposes very severe constraints on the telecentricity requirements. Fig~\ref{fig:telecentricity} shows the distribution of the theoretical telecentricity (in degree) considering 3~specific fields over all channels: the centre and the 2~extremes. It appears that all fields are comprised in the range $\pm$0.034$^\circ$. However, considering optical manufacturing defects and aligment errors, the telecentricty of the overall system is about $\pm 0.1^\circ$ in the worst case.

%%%%% FIGURE 5 %%%%%
%%%%%  

%**************** TABLE REQUIREMENTS **********************
\begin{table*}
\caption{\label{tab:requirements} Main IFU requirements.} 
\begin{ruledtabular}
\begin{tabular}{rrl}
 			&& $\bullet$\; The field of view (FOV), defined by the region of the sky imaged on the IFU slicer\\
Entrance FOV	&& \ \ \ \  stack, shall be at least 1.2mm$\times$1.2mm.\\
			%\cline{3-3}
			&& $\bullet$\; The IFU shall accomodate a geometric aperture of F/12.5 at the IFU entrance field.\\
\hline
			&& $\bullet$\; The IFU's virtual slit shall be divided into two sets of 15 sub-slits each.\\
			%\cline{3-3}
			&& $\bullet$\; The IFU exit focal plane (the focal plane of the IFU virtual slit) shall \\
Exit Focal Plane && \ \ \ \ nominally coincide with the entrance IFU focal plane.\\
			%\cline{3-3}
Virtual Slit	&& $\bullet$\; The IFU virtual slit shall be telecentric to the IFU field to within an angle of 0.2$^o$.\\
			%\cline{3-3}
			&& $\bullet$\; The IFU shall preserve the geometric aperture (F/12.5) at the IFU exit plane.\\
\hline
Spatial sampling && $\bullet$\; The average spatial sampling distance of the IFU along the spatial direction \\
			&& \ \ \ \ in the IFU exit image plane shall be 40~$\mu$m~$\pm$~1~\%.\\
\hline
Virtual slit dimensions	&& $\bullet$\; The average FWHM of each IFU virtual sub-slit along the spectral direction \\
			&&  \ \ \ \ in the IFU exit image plane shall be 80~$\mu$m~$\pm$~1~\%. \\
\hline
Wavelength band && $\bullet$\; The IFU shall meet the functional requirements over a wavelength range spanning \\
			&& \ \ \ \ from 0.7 to 5~$\mu$m.\\
\hline
Optical transmission && $\bullet$\; The average optical throughput shall be $\geq$50\% for any wavelength in the \\
			&& \ \ \ \ operating range and for any position within the FOV.\\
\hline
Image quality	&& $\bullet$\; The overall rms wavefront error (WFE) shall be less than 100~nm. \\
\hline
IFU envelope	&& $\bullet$\; The IFU shall fit within the static envelope: 190~$\times$~100~$\times$~60 mm. \\
\hline
Temperatures	&& $\bullet$\; All requirements given below apply at operating temperature and cryogenic \\
			&& \ \ \ \ environement (35-40~K).\\
\end{tabular}
\end{ruledtabular}
\end{table*}

%%%%%%%%%%%%%%%%%%%%%%%%%%%%%%%%%%%%%%%%%%%%%%%%%%%%%%%%%%%%%%%%
\section{Fan-shaped design versus classical design}  \label{sect:comparison}

\begin{figure}
   \begin{center}
   \begin{tabular}{c}
   \includegraphics[width=7cm]{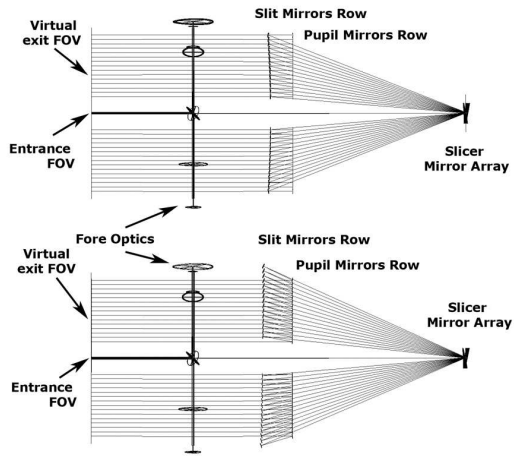}
   \end{tabular}
   \end{center}
   \caption
   { \label{fig:FAN} 
Overall layouts (seen from above) for both the classical image slicer (top) and the fan-shaped image slicer (bottom). The fore-optics used for the JWST/NIRSpec proposal is also presented. Considering the classical image slicer, pupil and slit mirrors are opposite one another and slit mirrors are disposed along a line. Considering the fan-shaped image slicer, pupil and slit mirrors are disposed in a fan-shaped configuration in order to minimize angles between incident and reflected beams on each elements. }
   \end{figure}

In order to estimate advantages of using the fan-shaped image slicer, we compared its performances with an image slicer using only classical principles briefly described in section~\ref{sect:intro} (see Fig.~\ref{fig:FAN}). Both image slicer are designed to cope with the NIRSpec/JWST requirements. It appears that all performances of the fan-shaped image slicer are better than those of the classical one and are all about the same level comparing all channels together.

%%%%% FIGURE 6 %%%%%
%%%%%  

%%%%% FIGURE 7 %%%%%
  \begin{figure}
   \begin{center}
   \begin{tabular}{c}
   \includegraphics[width=5cm]{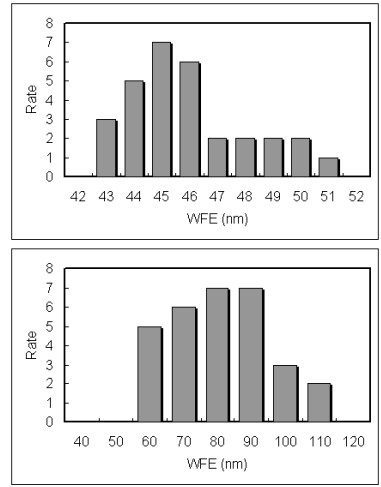}
   \end{tabular}
   \end{center}
   \caption
   { \label{fig:WFE} 
Theoretical wavefront error (T-WFE) distribution for the fan-shaped (top) and the classical image slicer (bottom). The dashed line marks the fore-optics mean T-WFE over the whole FOV.}
   \end{figure} 
%%%%%  

As an example in Fig.~\ref{fig:WFE}, we consider the distribution of the theoretical WFE (T-WFE) of both fan-shaped and classical image slicers. The T-WFE of the fan-shaped image slicer is about 45.5$^{+5}_{-3}$~nm and the T-WFE of the classical image slicer is about 76.1$^{+30}_{-25}$~nm. It is interesting to note that the fore-optics (see section~\ref{sect:design}) provides a mean T-WFE of about 45.7~nm over the whole FOV (marked by a dashed line in Fig.~\ref{fig:WFE}). Thus, the fan-shaped image slicer preserves the image quality of the fore-optics over the whole FOV (all channels) while the classical image slicer degrades the T-WFE by a factor two in certain channels. Furthermore, the distribution of the T-WFE of the fan-shaped image slicer is sharp since the range of values is about~8~nm over all channels allowing to guaranty a negligible level of differential aberration in the field.

%%%%%%%%%%%%%%%%%%%%%%%%%%%%%%%%%%%%%%%%%%%%%%%%%%%%%%%%%%%%%%%%
\section{Conclusion}

This article presents an original concept of image slicer called "Fan-shaped". Its design delivers good and homogeneous image quality over all IFU elements. We successfully apply its design to JWST/NIRSpec. Here we didn't discuss about manufacturing aspects since the performance aspects were preponderant however further investigations are under studying to drastically reduce costs and manufacturing aspects in such a design by preserving performances. Furthermore, a prototyping of the IFS (IFU and spectrograph) for the SNAP application is undergoing at LAM~\cite{Aumeunier2005}.

%\begin{acknowledgments}
%\end{acknowledgments}
%\appendix 
%\newpage %Just because of unusual number of tables stacked at end

\end{document}